\documentclass[]{iucr} 
\usepackage{graphicx,amsmath,braket,tikz,amssymb}
\journalcode{J}            
\begin{document}               

\title{Alignment facility and software for single crystal time of flight neutron spectroscopy}

\author[a]{Zihao}{Liu}
\author[a,b,c]{Harry}{Lane}
\author[c]{Christopher D.}{Frost}
\author[c]{Russell A.}{Ewings}
\author[b]{J. Paul}{Attfield}
\author[a]{Chris}{Stock}

\aff[a]{School of Physics and Astronomy, University of Edinburgh, Edinburgh EH9 3JZ, UK}
\aff[b]{School of Chemistry, University of Edinburgh, Edinburgh EH9 3FJ, UK}
\aff[c]{ISIS Pulsed Neutron and Muon Source, STFC Rutherford Appleton Laboratory, Harwell Campus, Didcot, Oxon, OX11 0QX, UK}

\maketitle 

\begin{abstract}
An instrument and software algorithm is described for the purpose of characterization of large single crystals at the Alignment Facility (ALF) of the ISIS spallation neutron source. We describe a method for both characterizing the quality of the sample and also aligning it in a particular scattering plane. We present a software package written for this instrument and demonstrate its utility by way of an example of the structural characterization of large singles crystals of Pb(Mg$_{1/3}$Nb$_{2/3}$)O$_{3}$. We suggest extensions and modifications of characterization instruments for future improved beamlines.   It is hoped that this software will be used by the neutron user community for pre characterizing large single crystals for spectroscopy experiments and that future facilities will include such a facility as part of the spectroscopy suite at spallation neutron sources.
\end{abstract}

\section{Introduction}

Modern developments of neutron chopper spectrometers have resulted in an unprecedented increase in neutron flux at the sample position along with an increase in angular resolved detector coverage.   Examples of such instrumentation exist at the ISIS neutron and muon spallation source and include MARI, MAPS~\cite{Ewings19:90}, MERLIN~\cite{Bewley06:385}, and LET~\cite{Bewley11:637}.  ISIS also hosts a series of backscattering instruments, OSIRIS and IRIS, providing exquisite resolution complementing these chopper spectrometers.~\cite{Telling05:7,Demmel18:1021}  All of these instruments have been extremely successful in contributing to solving problems in strongly correlated physics and hence are heavily over-subscribed.  Recently developed chopper spectrometers are particularly optimized for use with single crystal materials and the prior alignment and characterization of single crystal samples is crucial to their continued success and also for optimal use of facility time.

A key tool for alignment and characterization of single crystals has been Laue x-ray diffraction which has been applied for both characterization of large single crystals, and also mapping of grains.~\cite{Whitley15:48,Chung99:86,Ice09:60}  However, a significant problem with lab-based Laue x-rays is that x-rays penetrate matter on the order of microns, thus probing only the structure close to the surface. This can lead to misleading or even unhelpful results when characterizing samples for neutron scattering experiments, which have sample sizes on the order of centimeters, thus necessitating the use of neutrons, which interact with materials via nuclear forces and hence are strongly penetrating and probe the bulk structure.

The use of neutrons to characterize samples before use on modern chopper spectrometers is desirable as alignment and characterization can be time consuming, eating into valuable beamtime. To facilitate this, the excitations group at ISIS maintains a beamline allowing users to align and characterize single crystals beforehand. This is particularly important in experiments using extreme sample environments where alignment of single crystals in a well defined scattering plane is a necessity and setup and commissioning experiments involving such environments is also time consuming, adding to ``dead'' overhead time of any experiment. The Institut Laue-Langevin (ILL) has also recognized the importance of a pre-experiment characterization and alignment facility. The OrientExpress beamline~\cite{Ouladdiaf06:385}, equiped with a neutron Laue camera mounted in a backscattering geometry, is used routinely before diffraction and spectroscopy experiments.  We note that this emphasis on the importance of pre-characterization has resulted in the development of the larger Cyclops Laue diffractometer.~\cite{Ouladdiaf11:44}

We discuss a methodology of understanding the diffraction data from the ALF alignment facility at the ISIS spallation source.  The ALF facility has been in operation in some form for approximately 15 years, however has recently undergone an upgrade with improved goniometers and more detectors providing wider angular coverage.  These upgrades have occurred at the same time as improvements to the chopper instruments making the use of the facility important for users given the increased capabilities at ISIS.  As a step before any spectroscopy experiment, it is now important for users to have access to equipment and software that can fully characterize samples in an automated way and also be visualized.  

Motivated by this need, we develop a rotation matrix formalism to generate pole figure maps of Bragg peaks in terms of two spherical polar angles and further discuss software developed to visualize data from ALF.  We further provide demonstration results of a test sample on a relaxor ferroelectric PbMg$_{1/3}$Nb$_{2/3}$O$_{3}$ previously used in spectroscopy measurements.~\cite{Stock18:2} We note that this approach and software suite is motivated by the application of strain instrumentation and software on the E3 and L3 diffractometers at the now decommissioned NRU reactor (Chalk River, Canada).  These instruments and software were often used to co-align samples and fully characterize samples before time consuming experiments on the C5 and N5 triple-axis spectrometers and indeed the test sample used in this paper has been previously characterized on these diffractometers.

\section{ALF alignment instrument:}
\label{ALF alignment}
\begin{figure}
    \centering
    \includegraphics[width=0.75\textwidth]{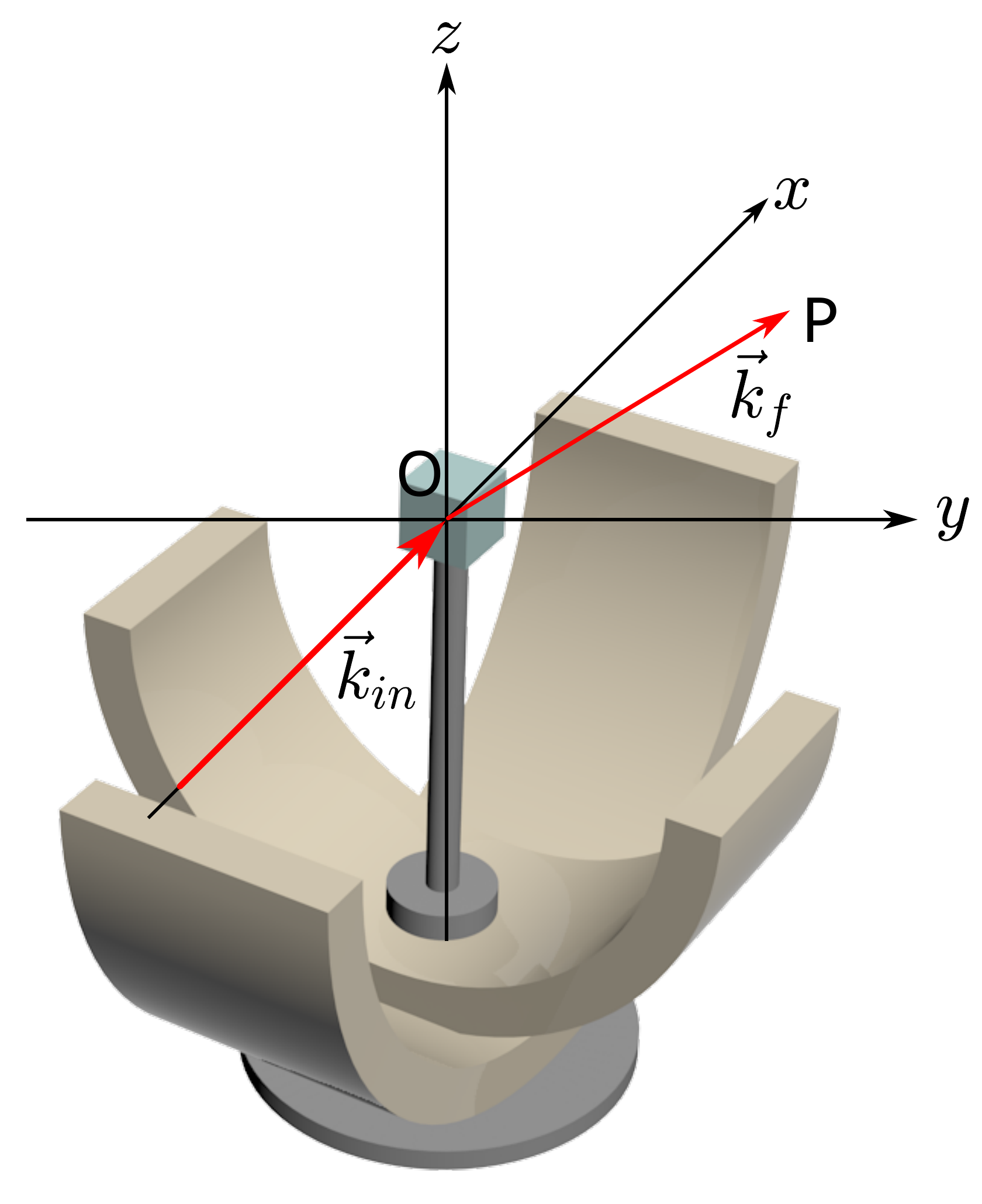}
    \caption{The layout of the goniometer and the definition of the lab frame. The red arrows show the scattering of an incoming neutron of wave vector $\vec{k}_{in}$ to a final neutron of $\vec{k}_{f}$ at pixel P.}
    \label{equipment}
\end{figure}

The alignment facility (ALF) has been constructed on port N2 of target station 1 at the ISIS spallation neutron source, viewing a liquid methane moderator. The instrument comprises supermirror guides from the source to 8.5 m of the total flight path, with gaps for ``t0" and disk choppers. These choppers produce a pseudo white beam of neutrons with wavelengths spanning the range from 0.2 to 5.2 \AA.  Thereafter the instrument flight path is composed of B4C collimation tube through to a sample position situated 14.86 m from the source. A bank of 37 position sensitive detector tubes of length 1 m is situated with its center point 1.32 m from the sample, covering a horizontal scattering angular range of 20.2$^{\circ}$ $<2\theta <$ 60.3$^{\circ}$. The layout of the current sample position at ALF is illustrated in Fig. $\ref{equipment}$. The sample position hosts a three-axis goniometer, with adjustable height and position parallel and perpendicular to the incident neutron beam. The lower goniometer rail is aligned along the $x$-axis and the upper is aligned at 90$^{\circ}$ along the $y$-axis.
These tilt rails are mounted on a rotation axis which rotates these axes around the $z$-axis. In front of the sample, a pseudo white beam of neutrons is incident on the sample with the resulting diffraction pattern on the detector being similar to an angular resolved Laue pattern. Data collection on ALF results in a time of flight image of the detector for a given goniometer setting.

We note that this instrument has no monochromatic chopper on the incident side and also no neutron wavelength analyzers (such as pyrolytic graphite crystals) on the scattered side.  In this work, we therefore assume that all measured  neutrons on the detector are elastically scattered, transfering no energy to the sample with the momentum transfer defined as $\vec{Q}=\vec{k}_{f} - \vec{k}_{in}$.  The elastic scattering condition implies that $E_{in}\equiv E_{f}$ implying $|\vec{k}_{in}|\equiv |\vec{k}_{f}|$.  The energy of the a given detected neutron then gives the magnitude of the scattered and incident wavevector via $E=\hbar^{2} k^2 /2m=\frac {1}{2} mv^{2}=\frac{1}{2} m (\frac{l}{t})^2$, where $t$ is the time of flight at the detector and $l$ the distance travelled from the target.  The magnitude of the momentum transfer is then given via Bragg's law $|\vec{Q}|=2|\vec{k}_{in}| \sin (\sigma)$.  The angle $\sigma$ is defined as half of the angle between $\vec{k}_{in}$ and $\vec{k}_{f}$. Under the assumption of elastic only scattering, the recorded time of flight of a neutron detected on a pixel of the position sensitive detector therefore defines the momentum transfer.

\section{Data visualization and the diffraction equations:}    

\begin{figure}
	\centering
	\includegraphics[width=0.75\textwidth]{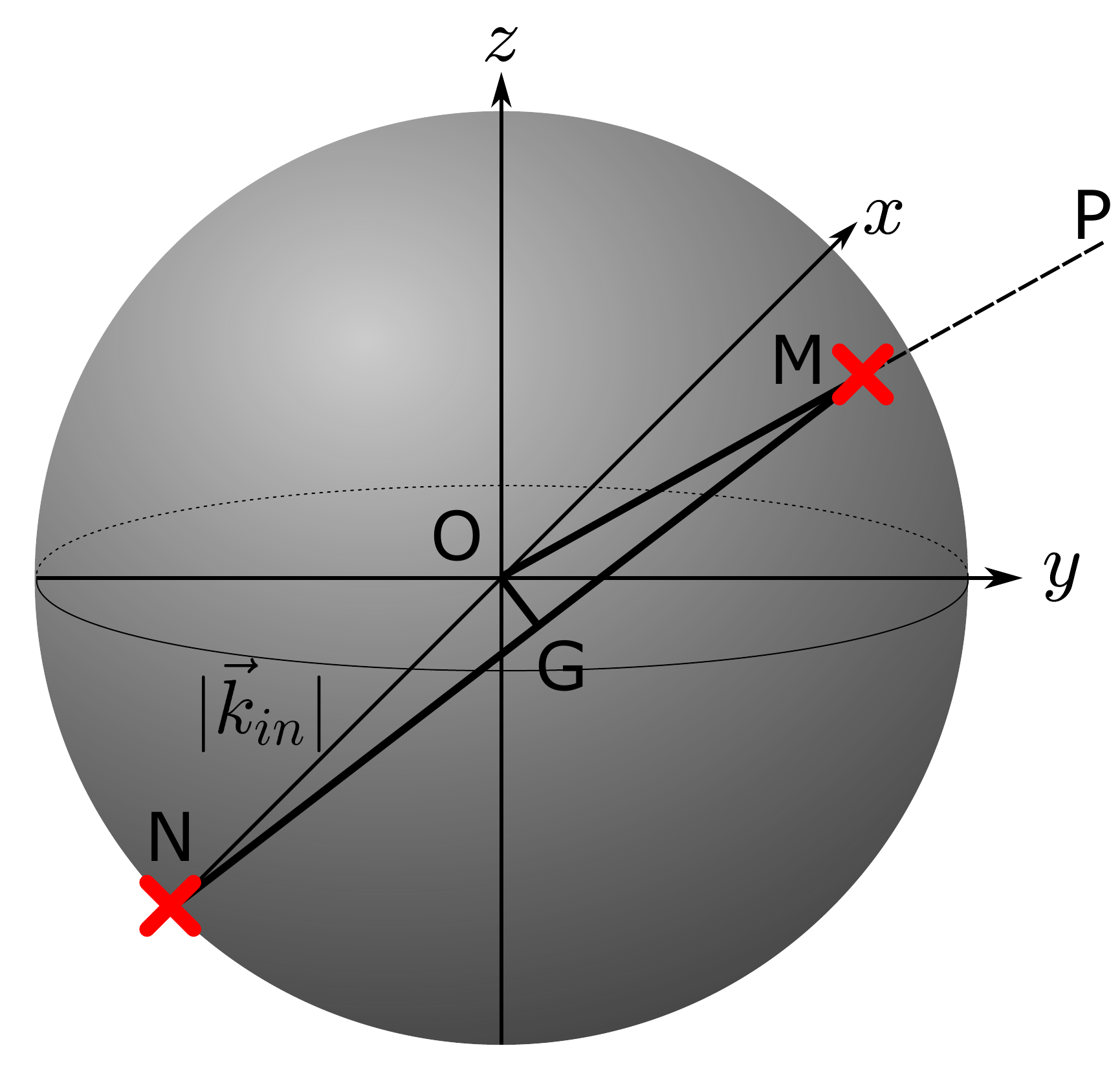}
	\caption{The Ewald sphere whose radius is $|\vec{k}_{in}|$ centered on the crystal position. $\protect\overrightarrow{OP}$ intersects the sphere at M and the x-axis intersects the sphere at N. G is a point on the sphere such that $\protect\overrightarrow{OG}$ bisects $\protect\overrightarrow{ON}$ and $\protect\overrightarrow{OM}$. }
	\label{imaginary}
	\end{figure}

Having stated the kinematics and the assumptions of ALF, we now discuss a methodology to visualize the data for a series of detector images taken at different goniometer settings. First, as shown in Fig. $\ref{imaginary}$, we consider the Ewald sphere of radius $|\vec{k}_{in}|$, centered on the crystal. The definition of $x$ and $y$ axes are the same as in Fig. $\ref{equipment}$. We denote the intersections of $\vec{k}_{in}$ and $\vec{k}_{f}$ with the sphere as the points N and M respectively. According to Bragg's law, the normal of the plane from which neutrons get diffracted bisects $-\vec{k}_{in}$ and $\vec{k}_{f}$. We can identify the midpoint of $\overrightarrow{NM}$ as the point $G$. Since $\overrightarrow{OG}$ bisects $\overrightarrow{ON}$ and $\overrightarrow{OM}$, we can use the location of G to indicate the orientation of the atom plane.

\begin{figure}
 	\centering
 	\includegraphics[width=0.75\textwidth]{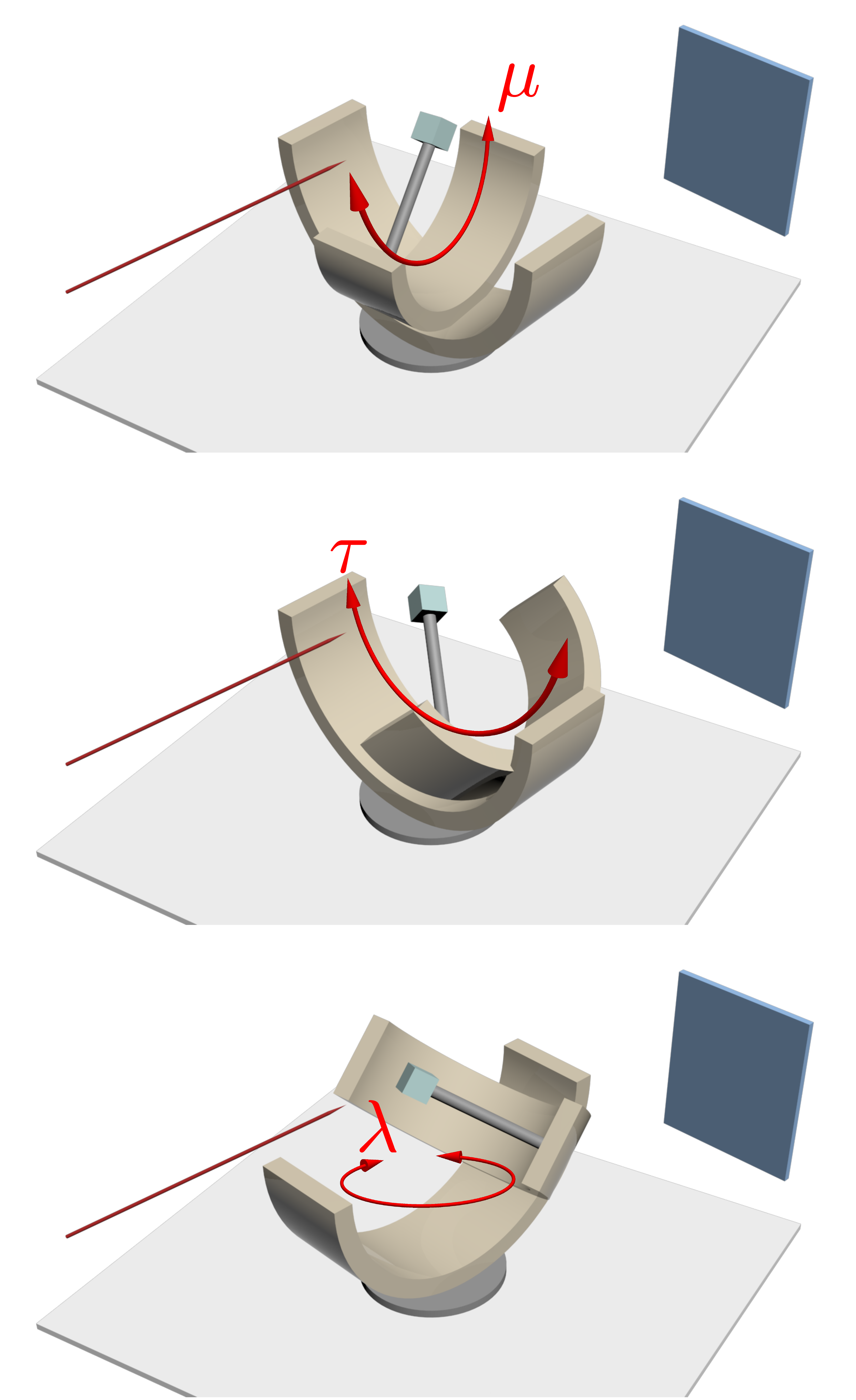}
 	\caption{Three frames indicating sequential movement of the goniometer about all three rotational axes. The thin blue cuboid represents the detector bank and the red arrow indicates the path of the incident neutron beam. Accessible sample rotation angles are limited by the angular range of the two goniometer rails.}
 	\label{randommove}
 	\end{figure}

Since $\overrightarrow{OG}$ is always perpendicular to the atomic plane from which the diffraction occurs, its relative location to the crystal is fixed. In other words, it rotates with the crystal. A rotation of the crystal about the three-axis relative to the lab frame (Fig. \ref{equipment}) can be described by three rotation matrices, representing the crystal's rotation around its base and along the lower and upper goniometer rails respectively.

	\begin{equation}
	\underline{G}_{rotated} = \underline{\underline{R}}_{rot} * \underline{\underline{R}}_{lower} * \underline{\underline{R}}_{upper} * \underline{G}_{lab}
	\label{rotupperlower} 
	\end{equation}
		
\noindent where $\underline{\underline{R}}_{rot}$, $\underline{\underline{R}}_{lower}$, and $\underline{\underline{R}}_{upper}$ are defined as

	\begin{equation}
	\underline{\underline{R}}_{rot} = {\begin{bmatrix}
		\cos(\lambda) & -\sin(\lambda) & 0\\
		\sin(\lambda) & \cos(\lambda) & 0\\
		0 & 0 & 1
		\end{bmatrix}}
	\end{equation}
	
	\begin{equation}
	\underline{\underline{R}}_{upper} = {\begin{bmatrix}
		1 & 0 & 0\\
		0 & \cos(\mu) & -\sin(\mu)\\
		0 & \sin(\mu) & \cos(\mu)
		\end{bmatrix}}
	\end{equation}
	
	\begin{equation}
	\underline{\underline{R}}_{lower} = {\begin{bmatrix}
		\cos(\tau) & 0 & -\sin(\tau)\\
		0 & 1 & 0\\
		\sin(\tau) & 0 & \cos(\tau)
		\end{bmatrix}}
	\end{equation}
	
\noindent and $\lambda$, $\mu$ and $\tau$ indicate the rotational angle of the crystal around the $z$, $x$ and $y$ axis, respectively. $\underline{G}_{lab}$ and $\underline{G}_{rotated}$ are column vectors describing the vector $\overrightarrow{OG}$, in Cartesian coordinates, of the lab and rotated frames respectively. One thing to note here is that $\underline{\underline{R}}_{rot}$, $\underline{\underline{R}}_{upper}$, and $\underline{\underline{R}}_{lower}$ do not commute with each other, hence their order matters. The order in which we applied those matrices in Eq. ($\ref{rotupperlower}$) can be justified by Fig. $\ref{randommove}$. It can be seen that any configuration of the crystal can be achieved by first moving the crystal along the upper rail (rotation around the $x$-axis) then the whole upper rail along the lower rail (rotation around the $y$-axis) and finally the base can be rotated to rotate the crystal, along with both goniometer rails, around the $z$-axis.
 
From each pixel of the detector, we can infer the location of $\underline{G}_{rotated}$ by geometry (Fig. \ref{imaginary}). We can then rotate this vector onto the lab frame by systematically inverting the rotation matrices defined in Eq. \ref{rotupperlower}. Application of this procedure to all files allows those with different gonimeter settings to be mapped onto the same coordinate system.


To begin with, we first need to show how the Cartesian coordinates of $\underline{G}_{rotated}$ can be deduced from each pixel's location. The pixels' coordinates, P, are defined by the three parameters $(l,\alpha,\beta)$, where $l$ is the distance between the pixel and crystal, and $\alpha$ and $\beta$ are the rotational angles around the $x$ and $z$ axes respectively. The pixel location is thus defined as

		\begin{equation}
		\underline{P} = \underline{\underline{A}} * \underline{\underline{B}} * {\begin{bmatrix} 
			l \\ 0 \\ 0 
			\end{bmatrix}}
		\label{BAP}
		\end{equation}	

\noindent where 

		\begin{equation}
		\underline{\underline{A}} = {\begin{bmatrix}
			1 & 0 & 0\\
			0 & \cos(\alpha) & -\sin(\alpha)\\
			0 & \sin(\alpha) & \cos(\alpha)
			\end{bmatrix}}
		\end{equation}
		\begin{equation}			
		\underline{\underline{B}} = {\begin{bmatrix}
			\cos(\beta) & -\sin(\beta) & 0\\
			\sin(\beta) & \cos(\beta) & 0\\
			0 & 0 & 1
			\end{bmatrix}}.					
		\label{AB}						
		\end{equation}

\noindent The Cartesian coordinates of $\underline{G}_{rotated}$ can be calculated from the geometric relation between itself and P.  Because $\overrightarrow{OG}$ and $\overrightarrow{OP}$ lie in the same plane, G can be mapped onto the $x-y$ plane by a rotation by angle $\alpha$, clockwise about the x-axis . Since $\overrightarrow{OG}$ bisects $\overrightarrow{ON}$ and $\overrightarrow{OM}$, a further anticlockwise rotation of angle $\frac{\beta}{2}$ about the $z$-axis aligns $\overrightarrow{OG}$ with the $y$-axis. 

That is to say

		\begin{equation}
		\underline{\underline{G}}_{rotated} = \underline{\underline{A}} * \underline{\underline{\Omega}} * {\begin{bmatrix} 
			0 \\ 1 \\ 0 
			\end{bmatrix}}
		\label{important2}
		\end{equation}	
		
\noindent where 

		\begin{equation}			
		\underline{\underline{\Omega}} = {\begin{bmatrix}
			\cos(\frac{\beta}{2}) & -\sin(\frac{\beta}{2}) & 0\\
			\sin(\frac{\beta}{2}) & \cos(\frac{\beta}{2}) & 0\\
			0 & 0 & 1
			\end{bmatrix}}.					
		\label{Aomega}						
		\end{equation}

\noindent Finally, combining Eq. (\ref{Aomega}) and Eq. (\ref{rotupperlower}), the locations of $\underline{G}_{lab}$ can be written as

  		\begin{equation}
		\underline{G}_{lab} = \underline{\underline{R}}_{upper}^{-1} * \underline{\underline{R}}_{lower}^{-1} * \underline{\underline{R}}_{rot}^{-1} * \underline{\underline{A}} * \underline{\underline{\Omega}} * {\begin{bmatrix} 
			0 \\ 1 \\ 0 
			\end{bmatrix}}.
		\label{reversematrix}
		\end{equation}


\noindent In defining the rotation matrices on the right hand side of eq. \ref{reversematrix}, five angles are needed, corresponding to the three goniometer angles and the two angles that define the pixel location. However, $\underline{G}_{lab}$ is uniquely defined by two spherical coordinates, $\theta$ and $\phi$, of the lab frame. Therefore, by scanning the five angles provided in each file, the resulting data can be visualized as a pole figure in the form of a two-dimensional colormap. We now illustrate the utility of this method by applying it to an exemplar dataset collected on ALF with a sample of single-crystal PMg$_{1/3}$Nb$_{2/3}$O$_{3}$.

\section{GUI description and test results:}  

We have applied the above formalism, based on rotation matrices, to design and construct a user interface for the ALF diffractometer. The aim of the program is not to control the instrument as has been implemented for triple-axis spectrometers~\cite{Lumsden06:385} and elsewhere for time-of-flight instruments,\cite{IBEX} but for visualizing data sets taken with different goniometer settings.  The goal of this program is to provide a way for a user to quickly visualize data systematically taken for a number of goniometer angle settings for the purposes of alignment and characterization.  To illustrate this software and the methodology, we have performed a test series of scans of a single crystal of PbMg$_{1/3}$Nb$_{2/3}$O$_{3}$ used in an experiment on MERLIN to map out the soft phonon modes with temperature.~\cite{Stock18:2}

The graphical user interface (GUI) written in Matlab has four tabs to perform different tasks needed for characterization and alignment.   The level of analysis achieved in each tab gradually increases and is based on the analysis in the previous tabs. 

\begin{figure}
	\centering
	\includegraphics[width=0.75\textwidth]{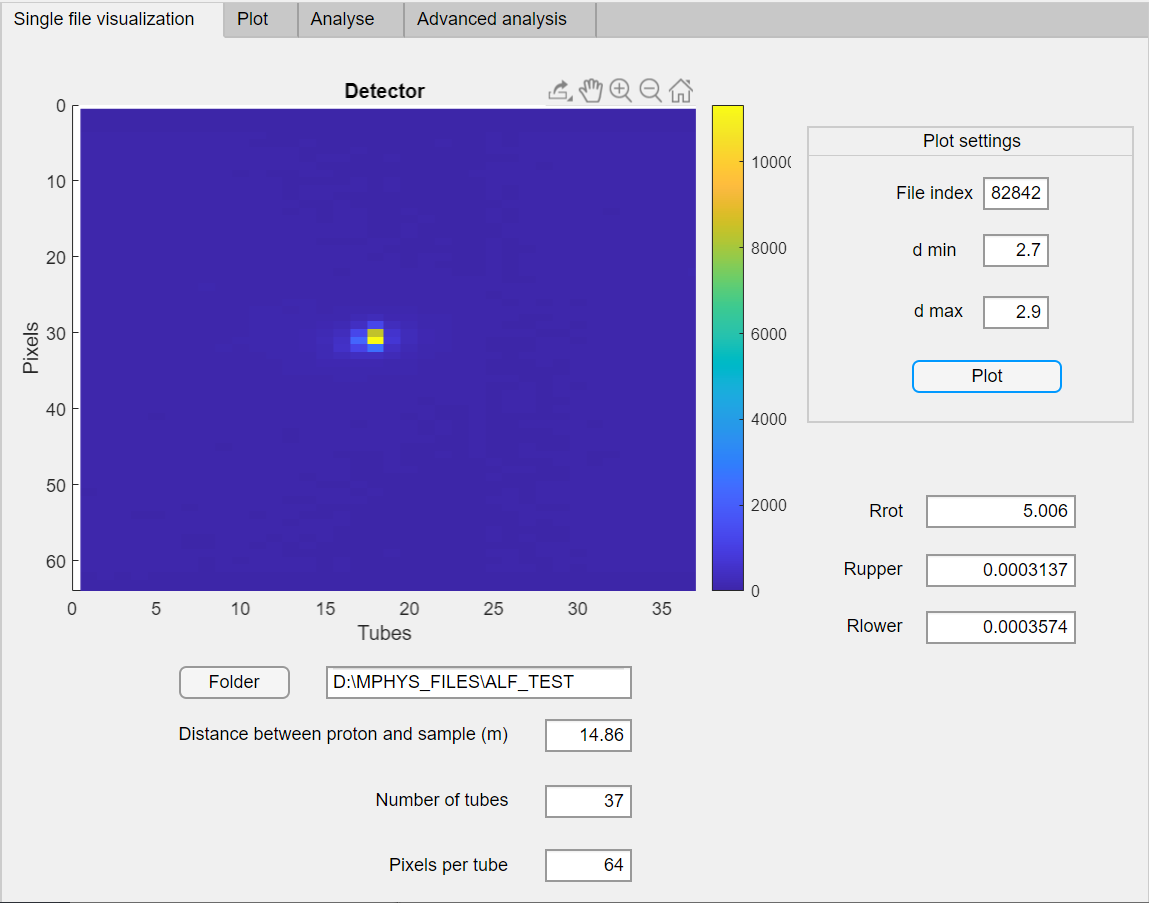}
	\caption{The first tab of the GUI which is used to visualize the detector from a single input file. The experiment parameters (distance between source and sample, number of tubes, pixels per tube) are auto-completed with the default values for ALF. However, these can be changed if future configurations require it (e.g. increased number of pixels per tube).}
	\label{firsttab}
	\end{figure}

\textit{Single file visualization:} The first tab (Fig. $\ref{firsttab}$) is used to visualize the detector bank as the experiment proceeds for a single angular setting of the goniometer illustrated in Fig. \ref{equipment}. The $x$-axis of the figure inside the first tab represents the detector tube index and the y-axis represents the pixel index within a given tube. When using this function, a user needs to input the file number and the $d$ spacing range for the Bragg peak defining the Miller planes of interest for visualization. The GUI will then plot the normalized count rate on the detector bank as a function of $x$ and $y$, and write the rotational angles $Rrot$ ($\lambda$), $Rupper$ ($\mu$), and $Rlower$ ($\tau$) defining the goniometer angles of the input file. 

\begin{figure}
	\centering
	\includegraphics[width=0.75\textwidth]{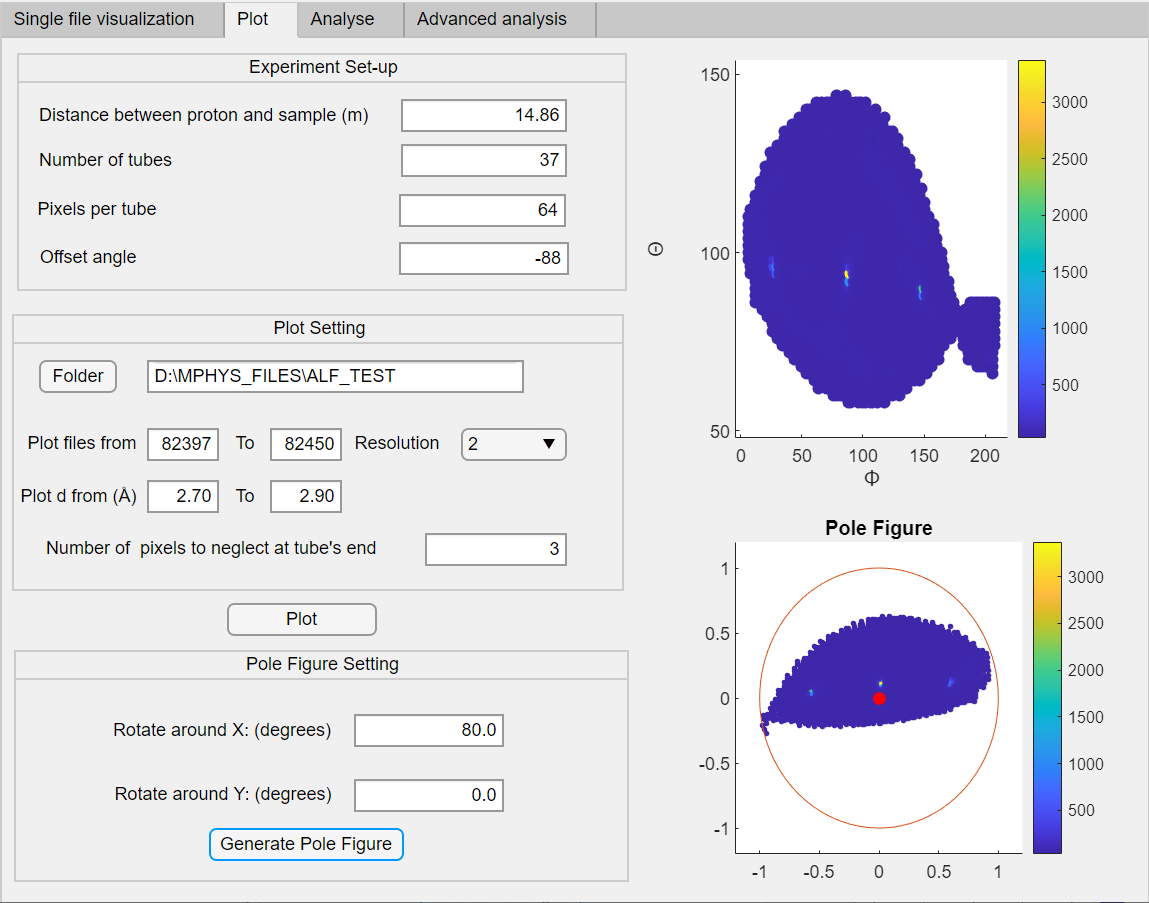}
	\caption{The second tab of the GUI to map all the peaks to the lab frame and generate a pole figure.}
	\label{secondtab}
	\end{figure}

\textit{Plot:} The second tab (Fig. $\ref{secondtab}$) enables the user to control how to map all the $\underline{G}_{rotated}$ files to the lab frame.  In the experiment set-up part, a user can determine whether to add an offset angle to $Rrot$. This step is necessary as in some experiments the two rails in Fig. $\ref{equipment}$ may not be initially aligned with the $x$ and $y$ axis used to define the rotation matrices in the section above.  Also, since the pixels at the extreme edges of the detector's tubes are subject to large read-out errors due to reduced efficiency, the user has the option to mask a certain number of pixels at each end of the tube to increase the quality of data and reduce contributing noise and error. The drop-down resolution menu lets the user choose the angular distance between two adjacent points in $\theta$ and $\phi$ to plot.  This defines the angular range over which the data is binned and can be tuned to increase plotted angular resolution, or to increase statistics through coarsening the angular resolution in the output plot. A user can therefore either choose a lower resolution to speed up the plotting process by coarsening the plotted angular steps or a higher setting for improved precision and visualization of grains or crystal mosaic. 

After mapping all of the $\underline{G}_{rotated}$ pixels to the lab frame,  the corresponding pole figure can also be generated. The pole figure pieces together all of the different detector images collected at different goniometer settings and plots them onto a common coordinate frame.  An Ewald sphere of fixed radius can be then be visualized in spherical coordinates based on two angles.   The projection of $\underline{G}_{rotated}$ on the equatorial plane is designed to be the intersection of the line connecting $\underline{G}_{rotated}$ and the south pole.  Therefore, if some points are in the southern hemisphere ($\theta > 90^{\circ}$) then all the points need to be mapped around the $x$ or $y$ axis for an angle determined by the user to avoid any loss of information and clarity in plotting. 
	
\textit{Analyse:} Then, in the third tab, some basic analysis can be performed. After loading the $\theta$ and $\phi$ plot generated in the second tab, a user can zoom in and click the figure to get the coordinates of a point. Those coordinates can be put into the GUI to calculate the angle between them. This step can be used to examine whether two peaks are of the same grain by comparing the angle with the theoretical prediction, and will be helpful for grain classification later in the fourth tab. 

In circumstances where the user thinks there is spurious point in the pole figure, they can input the range of $\theta$ and $\phi$ of the region of interest and the GUI will return a list of all the files involved which could be used for further debugging or re-measurement. 

	\begin{figure}
	\centering
	\includegraphics[width=0.75\textwidth]{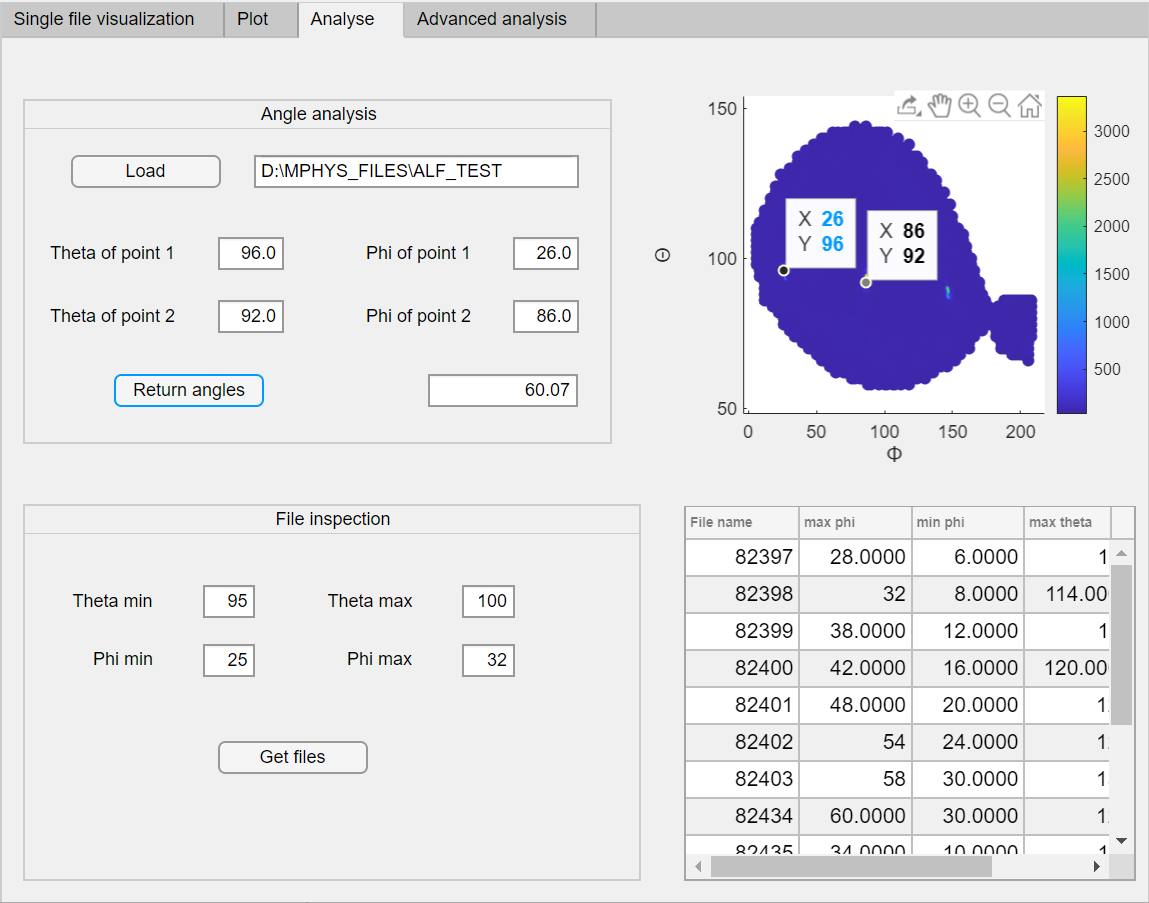}
	\caption{The third tab of the GUI. Angles between two peaks can be calculated and files containing points within the region of interest can be traced.}
	\label{thirdtab}
	\end{figure}

\textit{Advanced analysis:} The fourth tab is divided into two parts. The first is designed to estimate the mosaic angle of the crystal, defined to be the width of a Gaussian fit to the peak shape. After loading the plot generated in the second tab, the user can move the sliders to create two lines of constant $\theta$ and $\phi$, respectively.  Then a constant $\theta$ or $\phi$ cut can be plotted.  The origin is the intersection of the two lines and this figure changes in real time when the user moves the sliders, providing a rapid way to characterize the crystal quality and understand the pole figure. 

	\begin{figure}
	\centering
	\includegraphics[width=0.75\textwidth]{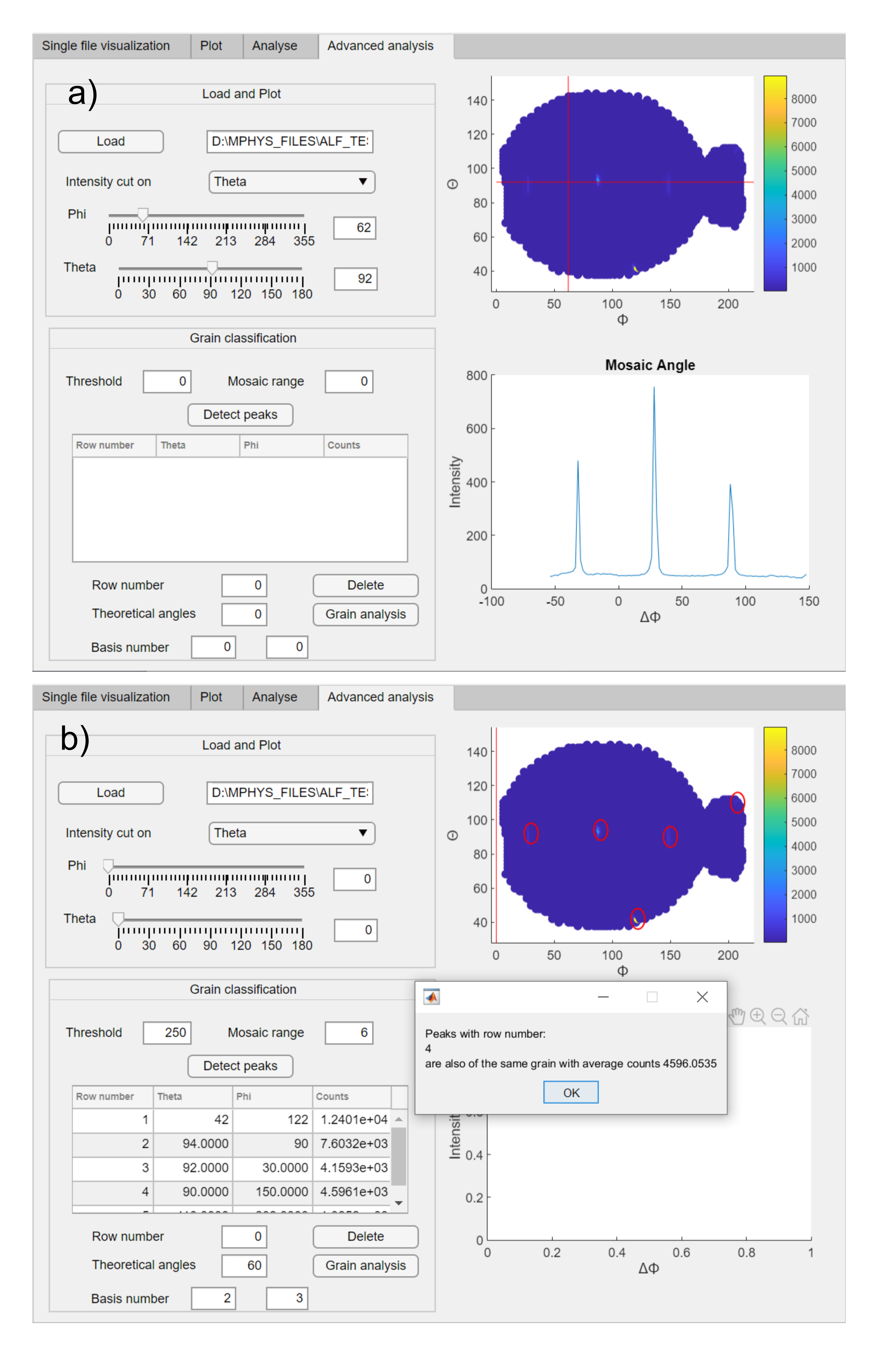}
    \caption{The fourth tab of the GUI. Part (a) is used to determine the mosaic angle of each peak along $\phi$ or $\theta$. Part (b) is a tool that allows for grain classification.}
 	\label{fourthtab}
    \end{figure}

The second part is used to determine how many grains a sample has and an estimate of the percentage of the total that each makes up. To do this, we first let the GUI detect and mark all the peaks with neutron counts higher than the user-defined threshold. Then, the user inputs a mosaic range so any point deviates from the peak's location within this range are considered to be part of the peak when summing the neutron counts. 

The GUI will then print a list of all the peaks detected with their locations in ($\theta$, $\phi$) and the corresponding integrated neutron counts. From this information, the user must select two peaks as bases that they know to be of the same grain. This step is necessary because at least two peaks are needed to uniquely define the location of a grain. Then, the user can input the theoretical angular separation of two adjacent peaks from the same grain and let the GUI search the list for peaks of the same grain. 

Once the GUI returns its search results the user will know which peaks in the plot are of the same grain and the average neutron counts of those peaks. The user can then take a note of the locations and the grain's average neutron counts before deleting them from the list. In cases where the neutron counts in a peak come from multiple grains, the user must neglect this peak and use the average neutron counts from other peaks of the same grain instead. This process can be done several times until all the peaks in the list have been identified. At the end of this process, the locations of each grain will be known and the ratio of the neutron counts from one type of grain to the total neutron counts detected can be taken as the approximate percentage of the prevalence of this grain in the crystal. 

\section{Conclusions}

The present paper has shown the utility of using ALF for characterization and subsequent alignment of large single crystals.  One thing that is evident from the pole figures constructed in this test experiment is that there is much angular range that is not accessible in the current configuration of ALF.  A means of improving the facility for future and more complete characterization of single crystals is the use of an Eularian cradle.  In this case, the extension of the software discussed here would be to replace the two tilt matrices listed above ($\underline{\underline{R}}_{lower}$,  $\underline{\underline{R}}_{upper}$) by a single rotation matrix and the inclusion of an inner rotation matrix.  The series of rotation matrices is described in Ref. \cite{Busing67:22}.    However, in the current instrumental configuration on ALF, through a combined scan of the tilt axes and rotation (Fig. \ref{equipment}) axes, much of the Ewald sphere can be measured to characterize the quality of the crystal and also to perform alignment for measurement on more advanced instruments.  We hope that by providing this methodology and outline that future excitations users will take advantage of this code and GUI, and that other spallation source facilities will support similar instruments to improve use of more complex chopper instruments.

\ack{Acknowledgements}
We thank the EPSRC and the STFC for funding and to J. A. Rodriguez and P. M. Gehring at NIST for helpful discussions.  H. L. was co-funded by the ISIS facility development studentship programme.  The authors are also grateful to the neutron scattering group at Chalk River, in particular Z. Tun, W. J. L. Buyers, and I. P. Swainson, for their help and support in many experiments at the NRU reactor that provided the motivation for this project.  The software package described here can be found at https://github.com/CSTOCK3/ALF-Analysis-Software.


\end{document}